\documentclass[fleqn,usenatbib]{mnras}
\usepackage{threeparttable}
\usepackage{booktabs}
\usepackage{newtxtext,newtxmath}

\usepackage[T1]{fontenc}

\DeclareRobustCommand{\VAN}[3]{#2}
\let\VANthebibliography\thebibliography
\def\thebibliography{\DeclareRobustCommand{\VAN}[3]{##3}\VANthebibliography}

\usepackage{graphicx}	
\usepackage{amsmath}	

\title[Star formation in interacting galaxies]{The Interaction Jigsaw: Investigating star formation in interacting galaxies}
{}

\author[Robin et al.]{
T. Robin$^{1}$\thanks{E-mail: robin.thomas@christuniversity.in},
Sreeja S Kartha$^{1}$\thanks{E-mail: sreeja.kartha@christuniversity.in},
Akhil Krishna R$^{1}$,
Ujjwal Krishnan$^{1}$,
Blesson Mathew$^{1}$,
T. B. Cysil$^{1}$, \and \hspace{0.1pt}
Narendra Nath Patra$^{2}$, and
B. Shridharan$^{1,3}$
\\
\\
$^{1}$Department of Physics and Electronics, CHRIST (Deemed to be University), Bangalore 560029, India\\
$^{2}$Department of Astronomy, Astrophysics and Space Engineering, Indian Institute of Technology Indore, 453552, India\\
$^{3}$Tata Institute of Fundamental Research, Homi Bhabha Road, Mumbai, 400005, India\\
}

\date{Accepted XXX. Received YYY; in original form ZZZ}

\pubyear{2024}

\begin{document}
\label{firstpage}
\pagerange{\pageref{firstpage}--\pageref{lastpage}}
\maketitle

\begin{abstract}

Interaction between galaxies play a pivotal role in their evolution. Ongoing star formation in spiral galaxies can be affected by these processes. Interacting galaxy pairs provide an opportunity to study these effects. We select a sample of interacting galaxies in field environments at various interaction stages and are nearly face-on and chose galaxy pairs NGC 2207/IC 2163, NGC 4017/4016 (ARP 305) and NGC 7753/7752 (ARP 86). We use the UltraViolet Imaging Telescope (UVIT) onboard \textit{AstroSat} to characterize the star-forming regions in the galaxy with a superior resolution of $\mathrm{\sim1.4\arcsec}$. We identified and characterized star-forming regions in the UVIT images of the sample and correlated them with the neutral hydrogen (H\textsc{i}) distribution. We detected localized regions of enhancement in star formation surface density ($\mathrm{\Sigma_{SFR}}$) and distortions in the sample of galaxies. We found this consistent with the distribution of H\textsc{i} in the galaxy. These are possible evidence of past and ongoing interactions affecting the star formation properties in the galaxies. We then conducted a study to understand whether the observed enhancements hold true for a wider sample of interacting galaxies. We observe a moderate enhancement in the star formation rate (SFR) with the interaction class, with the maximum of 1.8 being in the merger class of galaxies. We studied the SFR enhancement for the main galaxies in our sample as a function of pair mass ratio and pair separation. We observe a strong anti-correlation between the SFR enhancement and pair mass ratio and no linear correlation between the enhancement and pair separation. This suggests that the enhancement in interaction-induced star formation may be more strongly influenced by the pair mass ratios, rather than the pair separation. We also infer that the pair separation can possibly act as a limiting parameter for the SFR enhancement.

\end{abstract}




\section{Introduction}

Galactic interactions can significantly influence galaxies' morphology, gas distribution, and star-forming activity. Observations of interacting galaxies in the Local Universe reveal enhanced star formation compared to non-interacting galaxies and can act as one of the primary indicators for the occurrence of the interaction event. \citet{Larson1978ApJ...219...46L} and \citet{Kennicutt1990NASCP3098..269K} observed that the galaxies in the Atlas of Peculiar Galaxies \citep{Arp1966apg..book.....A} exhibited a wider range of optical colours and star formation rate (SFR) than typical galaxies. The observed increase in star formation activity has been studied in the infrared \citep{Kennicutt1987AJ.....93.1011K, Nikolic2004MNRAS.355..874N, Geller2006AJ....132.2243G, Lin2007ApJ...660L..51L}, optical \citep{Kennicutt1987AJ.....93.1011K, Donzelli1997ApJS..111..181D, Barton2000ApJ...530..660B, Barton2003ApJ...582..668B, Lambas2003MNRAS.346.1189L} and UV bands \citep{Knapen2015MNRAS.454.1742K}. 

However, the influence on galactic evolution due to interaction events is not well understood. Interacting galaxies experience physical processes such as tidal forces, gas shocks, and enhanced gravitational perturbations, which may affect the star formation properties of the galaxies \citep{Barnes1992ARA&A..30..705B,Keel1993AJ....106.1771K,Saitoh2009PASJ...61..481S}. The long-term sustainability of these effects on SFR remains uncertain, as it is unclear whether the observed alterations in the star formation activity persist over extended periods or is a transient phenomenon. Examining these questions is essential to understand the interaction-induced variations on galaxy evolution \citep{Montuori2010A&A...518A..56M}. 

The study of star formation activity can help us understand the galaxy's evolution by understanding the spatial distribution of young stars in the galaxy. Young massive stars of the O and B spectral types trace recent or ongoing star formation. These stars emit copious amounts of flux in the ultraviolet continuum \citep{Kennicutt1998}. This study aims to delve into the impact of these interactions on the star formation processes, seeking to elucidate the temporal evolution and spatial distribution of star-forming regions within the interacting systems. UV-bright regions, indicative of intense star formation, provide a unique observational window into the influence of environmental processes on star formation \citep{Kennicutt1998}. Studying the UV star formation properties of galaxies will thus help us understand the possible effects of interaction episodes in the current epoch. By focusing on UV-bright regions, we aim to understand regions within the interacting systems where the influence of the interaction is most pronounced. This facilitates a detailed examination of the physical conditions, such as the location of the region in the galaxy, gas density, etc., that contribute to the observed increase in ultraviolet emission, shedding light on the intricate relationship between star formation properties and galactic evolution.

\begin{figure*}
\begin{center}
{\includegraphics[width = \textwidth]{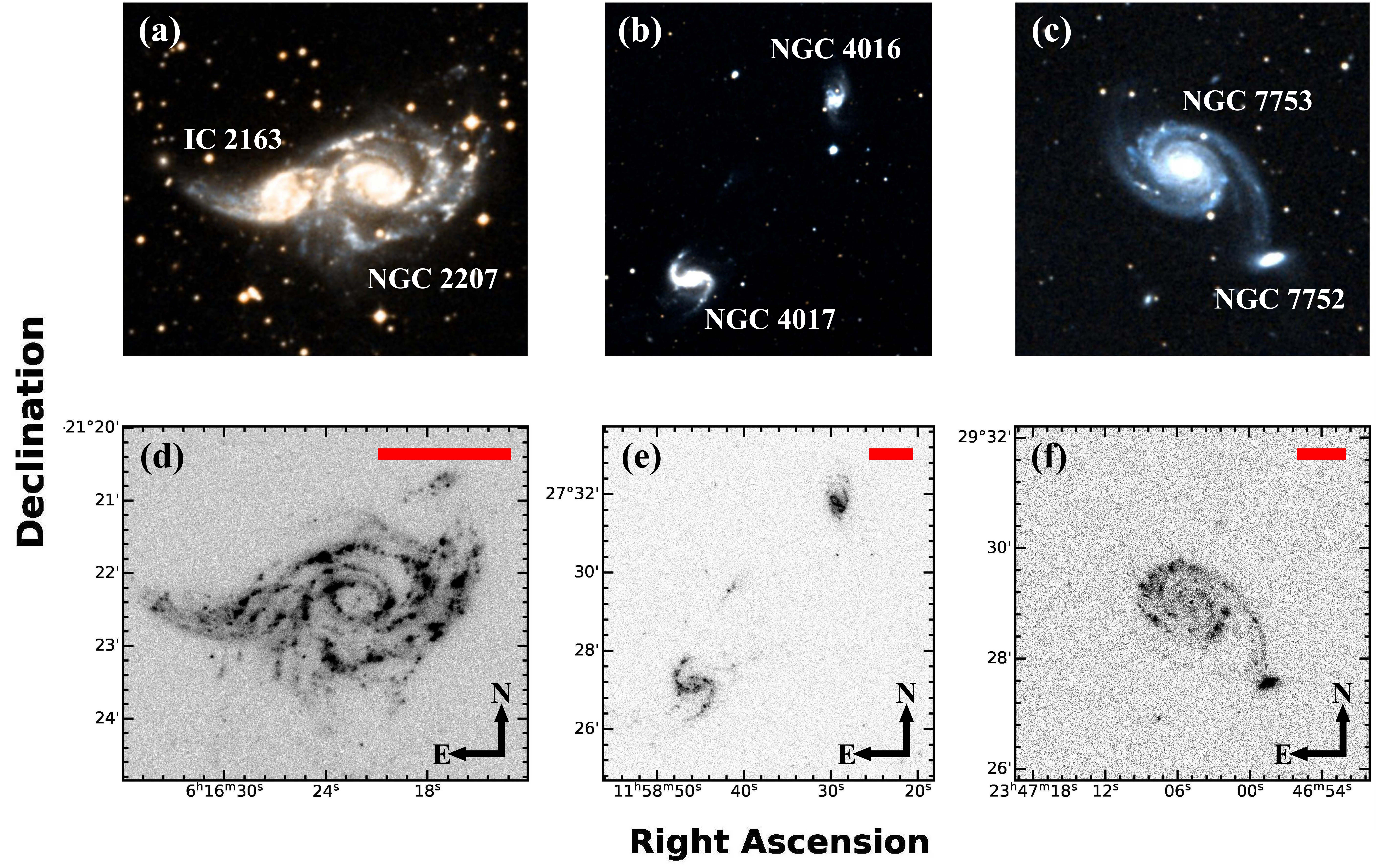}}

\caption{Colour composite images from the HiPS survey and corresponding UVIT image for the sample of galaxies are shown in the top and bottom panels, respectively. {The galaxy pairs in our sample are labelled in panels (a), (b) and (c). The emission in FUV1 for NGC 2207/IC 2163 and NGC 4016/4107 and in FUV2 for NGC 7752/7753 are shown in panels (d), (e) and (f), respectively. The red bars in the bottom panels represent an extent of 20 kpc at the distance to the galaxy pairs}. The foreground stars in the FOV were confirmed and removed using \textit{Gaia} DR3 catalogue \citep{Gaia2021A&A...649A...1G}. }
\label{fig:composite1}
\end{center}
\end{figure*}
We aim to carry out a comparative study on the star formation properties of UV-bright regions for a sample of interacting pairs of galaxies. We also try to understand and quantify the change in star formation processes due to the interaction event in the sample. Our study can help refine our understanding of the larger-scale processes that drive the evolution of the Universe. With the advent of UltraViolet Imaging Telescope (UVIT) and its superior spatial resolution of $\sim 1.4\arcsec$, we have an excellent opportunity to study the star formation properties on resolved scales. For our study, we present the UVIT Far-UV (FUV) observations of interacting galaxy pairs NGC 4017/4016 (ARP 305), NGC 7753/7752 (ARP 86) and NGC 2207/IC 2163. We aim to leverage the high-resolution UV data obtained from the UVIT to gain insights into the evolution of the selected sample of galaxies. As a pilot study, we studied the effects of interaction events on the star formation properties and the evolution of NGC 1512 in \citet{Robin2023arXiv230815537R} (henceforth, Paper I). This study's findings affect our broader understanding of galaxy evolution and contribute observational constraints to theoretical models. By quantifying the changes in star formation rates and identifying the spatial patterns of UV-bright regions, we seek to link the initial stages of interaction-induced star formation to the overarching evolutionary sequence of galaxies in the Universe.

We also investigate the distribution of neutral hydrogen (H\textsc{i}) in correlation with the UV star-forming activity. Galactic interactions often trigger the redistribution of gas. The distribution of H\textsc{i} tends to extend beyond the stellar radius of a galaxy \citep{Broeils1997A&A...324..877B} and can act as a reservoir for future episodes of star formation \citep{Kauffmann2015MNRAS.450..618K}. By correlating the UV-bright regions with the distribution of H\textsc{i}, we aim to disentangle the contribution of H\textsc{i} to the observed changes in star formation rates. 

We have adopted a flat universe cosmology throughout this paper with $\mathrm{H_0 = 71\ kms^{-1} Mpc^{-1} }$ and $\mathrm{\Omega_M = 0.27}$ \citep{Komatsu2011ApJS..192...18K}. The order of the paper is as follows. Section \ref{sample} briefly describes the selected sample of interacting galaxies, followed by a discussion on the data and analysis in Section \ref{Data}. We highlight the obtained results in Section \ref{results}. The summary is given in Section \ref{discussion}.

\begin{table*}
  \centering

  \caption{Basic parameters of selected galaxy pairs}
   \begin{tabular}{lcccccc} 
  \hline

  Parameter & NGC 2207 & IC 2163 & NGC 4016 & NGC 4017 & NGC 7752 & NGC 7753 \\
  \hline
  Type   & SAB(rs)bc pec$^{1}$ & SB(rs)c pec$^{1}$  & SBdm: $^{1}$  &  SABbc \tnote{1} &  I0:$^{1}$  &  SAB(rs)bc$^{1}$  \\
  RA [hh mm ss] &06 16 22.030$^{2}$ & 06 16 27.980$^{2}$ & 11 58 29.020   $^{2}$  &  11 58 45.671  $^{2}$ & 23 46 58.56$^{2}$ & 23 47 04.83$^{2}$ \\
  Dec [dd mm ss]  & -21 22 21.60$^{2}$ & -21 22 33.10$^{2}$ & +27 31 43.62  $^{2}$& +27 27 08.79 $^{2}$& +29 27 32.1$^{2}$ & +29 29 00.4$^{2}$  \\

  Distance (Mpc) &35$^{3}$ & 35$^{3}$ & 55  $^{4}$&  55 $^{4}$ & 68$^{5}$ & 68$^{5}$ \\
  log(Stellar Mass [$\mathrm{M_{\bigodot}}$]) & 10.98$^{6}$&10.77$^{6}$  & 9.21 $^{7}$& 9.89 $^{7}$ &9.85$^{8}$ &11.18$^{9}$ \\
  \hline
  \end{tabular}
\begin{tablenotes}
    \item $^{1}$ \citet{deVaucouleurs1991},  
    $^{2}$ \citet{JarrettRADEC2000AJ....119.2498J},
    $^{3}$ \citet{Elmegreen1995ApJ...453..100E},
    $^{4}$ \citet{Sengupta2017MNRAS.469.3629S},
    $^{5}$ \citet{Sengupta2009MNRAS.397..548S},
    $^{6}$ \citet{2207_Mass},
    $^{7}$ \citet{4017_Mass},
    $^{8}$ \citet{7752_Mass},
    $^{9}$ \citet{7753_Mass}

\end{tablenotes}
\label{tab:parameter}

\end{table*}

\section{Sample Selection}\label{sample}
Paper I conducted a pilot study focused on comprehending the impact of interactions on the evolutionary dynamics of the interacting galaxy pair NGC 1512/1510. Building upon this groundwork, the current study explores the intricacies of star formation properties exhibited by interacting pairs of galaxies within field environments. We identify interacting galaxy pairs in the field environments within the 100 Mpc having UVIT observations. Subsequently, we select interacting galaxy pairs at various stages of interaction in terms of the time of their closest approach. Our selection criteria further refine the sample, limiting it to pairs that are nearly face-on (inclination angle, \textit{i}, less than 40$^{\circ}$). Employing these criteria, we identify three interacting galaxy pairs meeting the specified conditions and possessing UVIT observations. The three galaxies selected for the study were NGC 2207/IC 2163, ARP 305 (NGC 4016/4017) and NGC 7753/7752 (ARP 86). In the following sections, we discuss these galaxy pairs briefly. These galaxies also have good quality H\textsc{i} 21 cm line observations from Very Large Array (VLA) and Giant Metrewave Radio Telescope (GMRT), which will be utilised in this work.

\subsection{NGC 2207/ IC 2163}

The spiral galaxies NGC 2207 and IC 2163 are currently involved in a near-grazing encounter \citep{Elmegreen1995ApJ...453..100E,ElmegreenA1995ApJ...453..139E}. The pair lies at a distance of 35 Mpc \citep{Elmegreen1995ApJ...453..100E}. NGC 2207 is classified as SAB(rs)bc pec while IC 2163 is classified as an SB(rs)c pec \citep{deVaucouleurs1991}. Tidal forces distorted IC 2163 in the in-plane direction, forming a tidal arm with two velocity components on the anti-companion side and forming an intrinsically oval disk with an eye-shaped ("ocular") morphology. Streaming motions in the oval are more than 100 km s$^{-1}$. Tidal forces also distorted NGC 2207 perpendicular to the plane, forming a strong, twisting warp. The H\textsc{i} in both galaxies has a large velocity dispersion of 30 - 50 km s$^{-1}$ and is concentrated in unusually large clouds (10$^{8}$ M${_\odot}$, \citealp{Elmegreen1993ApJ...412...90E}). Numerical simulations reproduced these peculiar structures and internal velocities with a prograde encounter affecting IC 2163. The time since the closest approach is 60 - 260 Myr \citep{Kaufman2012AJ....144..156K}.  We refer to NGC 2207 as the main galaxy and IC 2163 as the satellite.

\subsection{NGC 4016/4017 (ARP 305)}

NGC 4016 and NGC 4017 are an interacting pair of galaxies with a mass ratio of $\sim$ 1:3 and optical radial velocities of 3441$\pm$1 and 3449$\pm$2 km s$^{-1}$ \citep{Makarov2014A&A...570A..13M}, respectively. The pair lies at a distance of 55 Mpc \citep{Sengupta2017MNRAS.469.3629S}. NGC 4016 is classified as SBdm galaxy while NGC 4017 is classified as SABbc \citep{deVaucouleurs1991}. We refer to NGC 4017 as the main galaxy and NGC 4016 as the satellite galaxy, based on the mass ratio. At optical and UV wavelengths, the galaxy pair displays clear signatures of tidal interaction. These signatures include a figure of an eight-shaped inner disc in NGC 4016, enhanced spiral arms in NGC 4017, a tidal bridge remnant projected between the pair, and four tidal dwarf galaxy (TDG) candidates \citep{Sengupta2017MNRAS.469.3629S}. The time of closest approach between the galaxies is estimated to be $\sim$300 Myr \citep{Hancock2009AJ....137.4643H}.

\subsection{NGC 7752/7753 (ARP 86)}

This interacting galaxy pair consists of a grand-design barred spiral galaxy, NGC 7753, with a small companion, NGC 7752, towards the end of one of the spiral arms. A tidal bridge connects these interacting pairs of galaxies. NGC 7753 has been classified as SAB(rs)bc and NGC 7752 is classified as I0 galaxy \citep{deVaucouleurs1991}. The pair is located at a distance of 68 Mpc \citep{Sengupta2009MNRAS.397..548S}. \citet{Marcelin1987A&A...179..101M} have determined the heliocentric velocities of NGC 7753 and NGC 7752 to be 5160 and 4940 km s$^{-1}$ respectively. The galaxy pair underwent an initial perturbation 600 Myr ago and the recent upward crossing occurred 50 Myr ago \citep{Laurikainen1993ApJ...410..574L}. \textit{2MASX J23470758+2926531}, a compact dwarf galaxy lying to the South-East of NGC 7752, has been studied as a possible TDG associated with the ARP 86 system \citep{Sengupta2009MNRAS.397..548S, Zhou2014RAA....14.1393Z}. We refer to NGC 7753 as the main galaxy and NGC 7752 as the satellite. \\

The selected sample consists of galaxy pairs that have undergone interactions over different periods. The sample also varies with respect to the angle of approach, the mass ratios of the galaxies in the pair, their interaction velocities and gas densities. Table \ref{tab:parameter} gives a brief overview of the parameters for the selected sample of galaxies.

\section{Data and Analysis}\label{Data}
\subsection{UV Data}
To understand the star formation properties in the selected sample of galaxies and thus to comprehend the mechanisms influencing their evolution, we use UVIT data from the \textit{AstroSat} ISSDC archive \citep{Kumar2012SPIE.8443E..1NK}. UVIT has three bands - FUV (130-180 nm), Near UltraViolet (NUV, 200-300 nm) and VISible (VIS, 320-550nm), with the capability to observe simultaneously in all three bands. The instrument has a spatial resolution of  $\sim 1.4\arcsec$ and $ 1.2\arcsec$ for FUV and NUV filters, respectively, with a plate scale of $\mathrm{\sim 0.416 \arcsec pixel ^{-1}}$. NGC 2207/IC 2163 has been observed in the broadband filter F154W (FUV1, henceforth), NGC 4017/4016 in FUV1 and N242W (NUV, henceforth) and NGC 7753/7752 in F148W (FUV2, henceforth). Details of the observations are discussed in Table \ref{tab:obser}.

\begin{table*}
    \centering
    \caption{Log of observations}
    \begin{tabular}{llllllll}
        \toprule
      \addlinespace[1ex]
  Name of galaxy pair & PI &Obs. ID&  Filter & Peak $\lambda$ (\AA) & Date of Obs. & Exp. Time (s) \\
        
      \addlinespace[1ex]
                  \addlinespace[1ex]
                  \hline
    NGC 2207/IC 2163 & Paulsell&$\mathrm{A05\_169}$&FUV1&1541 &  12 December 2018& 11908.8\\
        \hline

      NGC 4017/4016 (ARP 305) & Kanak Saha &  $\mathrm{A04\_201}$ & FUV1 & 1541 &  13 January 2018 & 5962.6 \\
         & & & NUV & 2418 &13 January 2018 & 3016.8 \\
        \hline
            \addlinespace[1ex]
    NGC 7753/7752 (ARP 86) & Mousmi Das&$\mathrm{A05\_104}$&FUV2&1481 & 29 May 2019 & 4140.5\\
        \hline

    \end{tabular}
    \label{tab:obser}
\end{table*}

We used the software package CCDLab \citep{Postma_Leahy_2017} to reduce the Level 1 data. We followed the procedure outlined in Paper I for the reduction and calibration of UVIT images. For the FUV1, FUV2 and NUV filters, the adopted zero point magnitude values are 18.097, 17.771 and 19.763 mag, respectively \citep{Tandon2020AJ....159..158T}. Figure \ref{fig:composite1} represents the colour composite images generated by HiPS survey and corresponding UVIT image for the sample of galaxies.

\subsection{H\textsc{i}  data}\label{dataHI}

H\textsc{i} plays a crucial role in determining the star formation in a galaxy. It is the primary fuel for star formation. The star formation rate density correlates with the column density of the H\textsc{i} across all different types of galaxies, from dwarfs \citep{Roychowdhury2019A&A...631A.115R,Patra2016MNRAS.456.2467P} to large spirals. Further, the H\textsc{i} in galaxies extends much beyond the stellar disks \citep{kamphuis1992kinematics, broeils1994search,Koribalski_1512_2009,mihos2012extended, Patra2016AstBu..71..408P}. As such, H\textsc{i} is one of the primary tracers of galaxy dynamics. The H\textsc{i} can trace the interaction between pairs of galaxies much better than its optical counterparts. Hence, H\textsc{i} complements the UV observation to trace the effect of interaction on star formation. We use the archival data from the VLA and GMRT to obtain the H\textsc{i} maps for our target galaxy pairs. For NGC 2207/IC 2163 and NGC 4017/4016 (ARP 305), H\textsc{i} column density maps were available from \citet{Elmegreen2016ApJ...823...26E} and \citet{Sengupta2017MNRAS.469.3629S}, respectively. For NGC 7753/7752 (ARP 86), analyzed H\textsc{i} column density maps were unavailable. However, raw data from the GMRT archive was available (PI: Chandreyee Sengupta, date of observation: 02 May 2008). We analyzed this GMRT data to obtain the H\textsc{i} spectral cube and H\textsc{i} column density map. We used Astronomical Image Processing Software (AIPS, \citealp{AIPS1985daa..conf..195W}) to reduce the data. Bad data from dead antennas and data corrupted by radio frequency interference (RFI) were flagged. The data were calibrated for amplitude and phase using primary and secondary calibrators. Thus, a clean, flagged, calibrated visibility cube is produced for further processing. The visibility cube is then imaged using AIPS task \textit{IMAGR} to produce a dirty image cube. We identify the channels containing line emission by visually inspecting the dirty image cube, and consequently, the AIPS task \textit{IMLIN} was used to subtract the continuum by building a continuum model using line-free channels. Further, a clean spectral cube is produced by cleaning the continuum-subtracted image cube using the AIPS task \textit{APCLN}. We use the AIPS task \textit{MOMNT} to produce the moment maps, i.e., the total intensity (moment zero) map, the velocity map (moment one), and the velocity dispersion map (moment two). We use the moment zero map to investigate the correlation between gas distribution and star formation in our interacting pair of galaxies. We note that the H\textsc{i} in interacting galaxies is diffused, and a map  of resolution $\sim 30 ^{\prime \prime}$ is utilized to trace the interaction and dynamics.

\subsection{Identification of star-forming regions and extinction}\label{Extinction}
We utilized the ProFound package \citep{Robotham_Davies_Driver_Koushan_Taranu_Casura_Liske_2018} to identify the brightest regions from the UVIT FUV images. The source finding and photometry details have been discussed in \citet{Ujjwal2022MNRAS.tmp.2176U} and Paper I. ProFound identified 194 UV bright regions in NGC 2207/IC 2163, 104 UV bright regions in NGC 4017/4016 and 61 UV bright regions in NGC 7753/7752 pair. We removed star-forming regions with FUV magnitude, $\mathrm{m_{FUV}}$, greater than 21 mag to exclude regions with greater photometric error ($\gtrsim$ 0.1 mag) from further analysis (\citealp{Ujjwal2022MNRAS.tmp.2176U}, Paper I, \citealp{Ujjwal2024A&A...684A..71U}). A final sample of 140, 70 and 40 bright star-forming regions was obtained for NGC 2207/IC 2163, NGC 4017/4016 and NGC 7753/7752, respectively. 

While studying star formation in the UV continuum, the major drawback is its sensitivity to extinction \citep{Kennicutt1998}. We take the rest-frame extinction into account for our study. For galaxy pairs NGC 4017/4016 and NGC 7753/7752, the internal extinction was estimated by obtaining the values of reddening coefficients, $\mathrm{c(H\beta)_{internal}}$ from \citet{Zasov2018MNRAS.477.4908Z} and \citet{Zhou2014RAA....14.1393Z}, respectively. We cross-matched the star-forming regions identified in these papers with those identified in this study for extracting the reddening coefficient values. In case of direct match, we adopted the coefficient values for the identified regions.  For the regions identified newly in our study, we assumed an average value of the reddening coefficient of the five nearest regions identified in the previous studies. In the case of NGC 2207/IC 2163 pair, we utilized the MIPS 24$\mu$m. Even though there exists a resolution mismatch between MIPS and UVIT images, MIPS $\mathrm{24 \mu m}$ is the best available dataset to account for internal extinction. To estimate the Galactic extinction, we adopted the $\mathrm{A_{v} =  0.238,\, 0.064,\, and\, 0.271\,}$ for NGC 2207/IC 2163, NGC 4017/4016 and NGC 7753/7752, respectively, as estimated in \citet{Schlafly2011}. We utilized the \citet{Cardelli1989ApJ...345..245C} extinction law, with $\mathrm{R_v = (A_V)/ E(B-V)} = 3.1$ and the Astropy extinction module \citep{Astropy2013A&A...558A..33A} for estimation of Galactic extinction. The total extinction correction for NGC 4017/4016 and NGC 7753/7752 was carried out using the discussion provided in Paper I, and for NGC 2207/IC 2163, the correction was carried out using the discussion given in \citet{Ujjwal2022MNRAS.tmp.2176U}. The corrected FUV magnitudes are used in further analysis.

\begin{figure*}
\begin{center}
{\includegraphics[width = 1\textwidth]{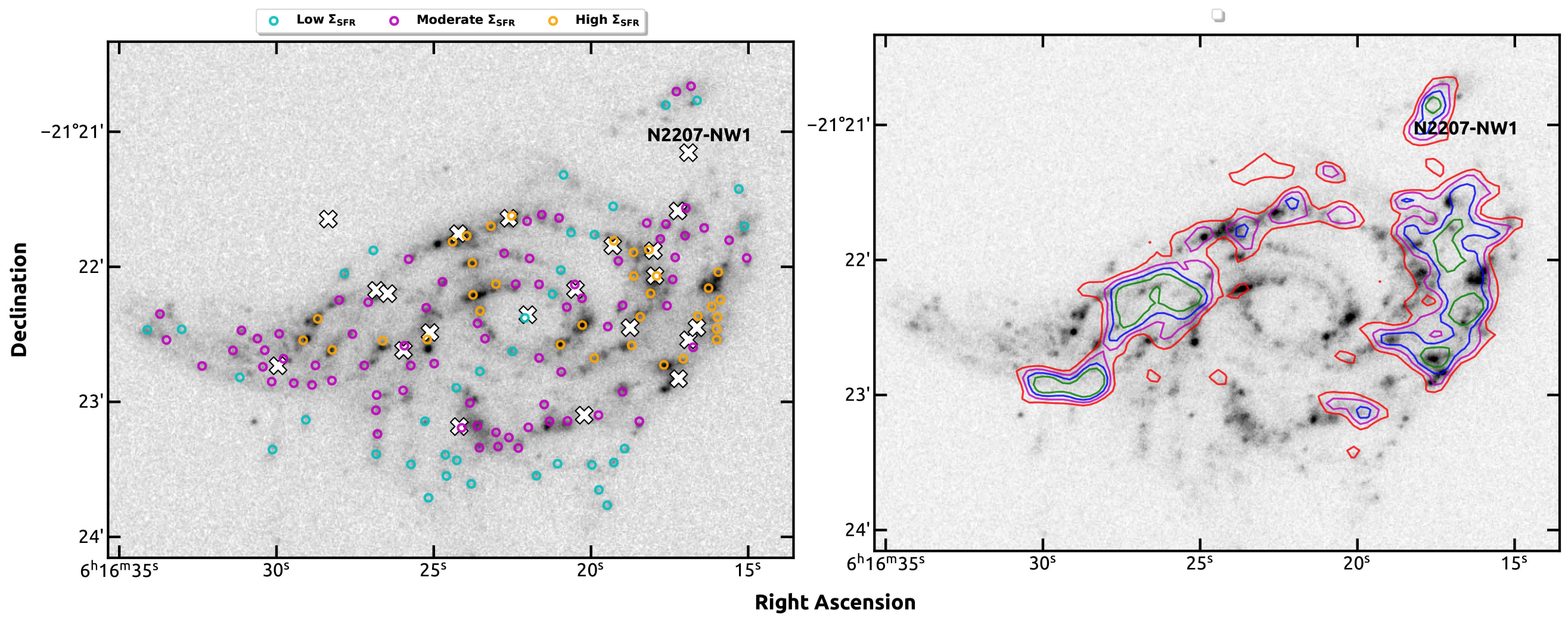}}
\caption{(Left) Spatial distribution of the $\mathrm{\Sigma_{SFR}}$ in the galaxy pair NGC 2207/IC 2163 plotted on the FUV image. We observe a non-localized enhancement spread out along the galaxy's spiral arms. Candidate \textit{N2207-NW1} is observed to host four star-forming regions, with a $\mathrm{log({\Sigma_{SFR}}_{mean} )}$ of $\sim -2.6$. { The white markers denote the the locations of the X-ray sources detected in \citet{Mineo2013ApJ...771..133M}.} (Right) The H\textsc{i} column density contours are overplotted on the FUV image of NGC 2207/IC 2163. The contours of red, magenta, blue and green colours represent H\textsc{i} column density values $\mathrm{(2.5, 3, 3.5, 4) \times 10^{21} cm^{-2}}$. Candidate \textit{N2207-NW1} might be in the process of forming a TDG.}
\label{fig:2207ssfr}
\end{center}
\end{figure*}

\section{Results and Discussion}\label{results}
\subsection{Estimation of star formation rates}\label{identity}

We acquired the number of pixels containing 100\% of the flux and estimated the area of each region. The counts were integration time-weighted to estimate the counts per second (CPS). The resulting values were converted to magnitude units using the zero point conversion (ZP) technique, as discussed in \citet{Tandon2020AJ....159..158T}. We calculated the star formation rate (SFR)  using equation 4 in \citet{Kaisina2013} and is as reproduced below.
\begin{equation}
    \mathrm{log(SFR[M_{\odot}\, yr^{-1}]) = 2.78 - 0.4 \times m^{c}_{FUV} + 2 \times log\ D}
\end{equation}

where $\mathrm{m^{c}_{FUV}}$ denotes the magnitude corrected for Milky Way and internal extinction and D is the distance to the galaxy in Mpc.

We studied the radial distribution of star formation to understand the influence of various galactic properties on star formation across the galaxy. We utilized the values of distance to the galaxy, position of the galactic center, inclination, and position angle, as given in Table \ref{tab:parameter}. We estimated the galactocentric distance in kpc for each region, using the discussion in \citet{Marel2001AJ....122.1807V}. We estimated the star formation rate surface densities ($\mathrm{\Sigma_{SFR}}$) of identified regions by dividing the estimated SFRs by the area of each region (in kpc$^2$). We will discuss the distribution of $\mathrm{\Sigma_{SFR}}$ along the galactic morphology for individual galaxy pairs in the following subsections.

\subsection{Studying the distribution of $\mathrm{\Sigma_{SFR}}$ along the galaxy pairs}\label{sigma}

We study the distribution of $\mathrm{\Sigma_{SFR}}$ across the galaxy pair to comprehend the intricate interplay between galaxy interactions and the regulation of star formation processes within galactic environments. By probing the spatial distribution of $\mathrm{\Sigma_{SFR}}$ across galaxy pairs, we can unravel the complex mechanisms influencing the modulation of star formation activity due to the interaction event. To study the magnitude of star formation enhancement across the selected sample of galaxies, we divided $\mathrm{\Sigma_{SFR}}$ into quartiles. Regions with $\mathrm{\Sigma_{SFR}}$ falling above the upper quartile are identified as high $\mathrm{\Sigma_{SFR}}$ regions. Regions with $\mathrm{\Sigma_{SFR}}$ below the lower quartile are identified as low $\mathrm{\Sigma_{SFR}}$ regions and Regions with $\mathrm{\Sigma_{SFR}}$ between the upper and lower quartiles are identified as moderate $\mathrm{\Sigma_{SFR}}$ regions. The range of $\mathrm{\Sigma_{SFR}}$ and the values of upper and lower quartiles for each galaxy in the selected sample is given in Figure \ref{fig:append}.

\subsubsection*{NGC 2207/IC 2163}

The distribution of $\mathrm{\Sigma_{SFR}}$ in the galaxy pair NGC 2207/IC 2163 is depicted in Figure \ref{fig:2207ssfr} (left panel). We observe that the identified star-forming regions exhibit $\mathrm{log({\Sigma_{SFR}})}$ in the range of -3.2 to -1.9. The galaxy pair is undergoing a near-gazing encounter \citep{Elmegreen2009IAUS..254..289E}. We observe that the region of interaction between the galaxies exhibits high $\mathrm{\Sigma_{SFR}}$ regions. The side of the main galaxy away from the satellite galaxy also exhibits high $\mathrm{\Sigma_{SFR}}$ regions. To understand the observed heightening of star formation in the region, we considered the distribution of X-ray sources throughout the galaxy pair from \citet{Mineo2013ApJ...771..133M}. {The distribution of these sources is represented by white "X" markers in Figure \ref{fig:2207ssfr} (left panel). We observe that the X-ray sources are hosted mostly in the northern part of the main galaxy. We also note a possible correlation between the number of X-ray sources with local $\mathrm{\Sigma_{SFR}}$. Taking into account the simulated trajectory of IC 2163 about NGC 2207 discussed in \citet{Struck2005MNRAS.364...69S}, we suggest that the observed enhancements in local $\mathrm{\Sigma_{SFR}}$ and X-ray sources, in addition to the disruption in morphology, are possibly due to the interaction between the galaxy pair. This supports our observations of high $\mathrm{\Sigma_{SFR}}$ in regions in the main galaxy away from the satellite galaxy.} We also observe a candidate to the North-West of the main galaxy, which has been identified as \textit{N2207-NW1}. Studies by \citet{ElmegreenA1995ApJ...453..139E, Elmegreen2009IAUS..254..289E} and \citet{ Elmegreen2016ApJ...823...26E} had observed this cloud in the galaxy system with no active star formation. We identify four distinct star-forming regions within this candidate, exhibiting a $\mathrm{log({\Sigma_{SFR}}_{mean})}$ of $\sim -2.6$.

\begin{figure*}
\begin{center}
{\includegraphics[width = 1\textwidth]{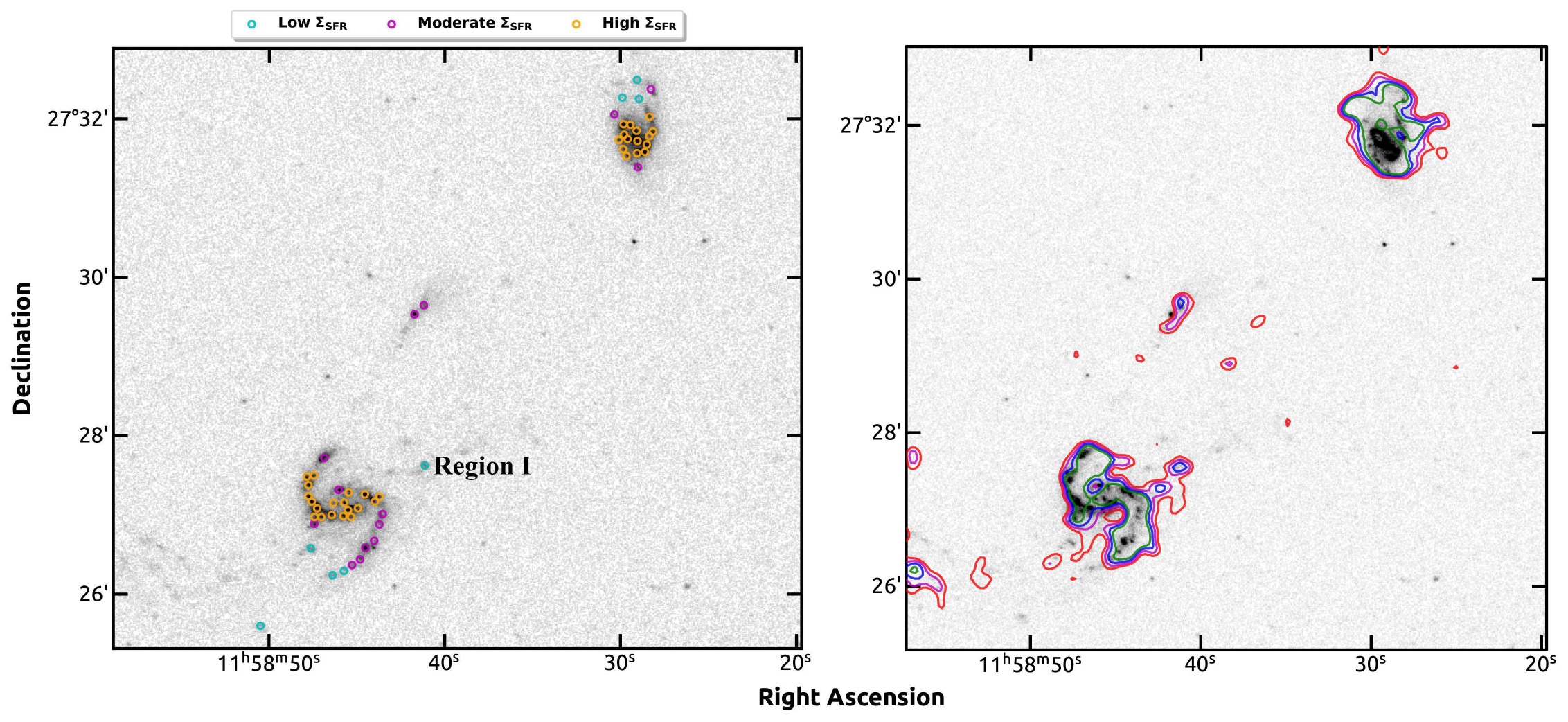}}
\caption{(Left)Spatial distribution of the $\mathrm{\Sigma_{SFR}}$ in the galaxy pair NGC 4017/4016 plotted on the FUV image. We observe moderate star formation in the main galaxy with $\mathrm{log({\Sigma_{SFR}})}$ in the range of -3.2 to -1.6. Moderate star formation is seen in different regions across the galaxy pair, which might have formed as a result of the interaction event. (Right)  The H\textsc{i} column density contours are overplotted on the FUV image of NGC 4017/4016. The contours of red, magenta, blue and green colours represent H\textsc{i} column density values $\mathrm{(0.8,1, 1.2, 1.4) \times 10^{21} cm^{-2}}$. }
\label{fig:305ssfr}
\end{center}
\end{figure*}

We overplotted the H\textsc{i} column density over the UV distribution as is shown in Figure \ref{fig:2207ssfr} (right panel). The contours were plotted for column density values of  $\mathrm{(2.5, 3,3.5,4) \times 10^{21} cm^{-2}}$. We observe that the contours trace the spiral arms of the main galaxy and some regions of the satellite galaxy. On analysing this distribution of H\textsc{i} with respect to the $\mathrm{\Sigma_{SFR}}$ distribution, we see that the regions in the spiral arms which show enhanced star formation activity correlate with the regions with higher H\textsc{i} column densities. However, this correlation is not true for all regions. We observed some regions with high column density but exhibit only moderate star formation ($\mathrm{log[{\Sigma_{SFR}}_{mean}] }$ < -2.4). It is possible that the high SFR was observed in H\textsc{i} dominated regions due to the large-scale synchronization of star formation in NGC 2207 where many of the giant molecular clouds formed in the outer parts are dispersed into H\textsc{i} before their OB associations have significantly faded \citep{Elmegreen2016ApJ...823...26E}. Candidate \textit{N2207-NW1} displayed elevated H\textsc{i} column density. Importantly, this candidate, falling within the same velocity range as the galactic pair NGC 2207/IC 2163, is deemed integral to the galaxy pair system. We suggest that \textit{N2207-NW1} may be in the early stages of forming a TDG. While previous studies by \citep{Struck2005MNRAS.364...69S, Kaufman2012AJ....144..156K,Lelli2015A&A...584A.113L} observed a gas cloud devoid of any star formation, our study has found ongoing star formation in \textit{N2207-NW1}. We report for the first time that there is ongoing star formation within the \textit{N2207-NW1}. This further supports the possibility of \textit{N2207-NW1} being a possible TDG candidate.

\subsubsection*{NGC 4017/4016}

The distribution of $\mathrm{\Sigma_{SFR}}$ in the galaxy pair NGC 4017/4016 is depicted in Figure \ref{fig:305ssfr} (left panel). The star-forming regions identified in the galaxy pair exhibit $\mathrm{log({\Sigma_{SFR}})}$ in the range of -3.2 to -1.6. We notice a moderate $\mathrm{\Sigma_{SFR}}$ in the spiral arms of the main galaxy, with $\mathrm{log({\Sigma_{SFR}})}$ ranging from -2.4 to -2. We also observe high $\mathrm{\Sigma_{SFR}}$ in the satellite galaxy, NGC 4016, in an eight-shaped ring structure. It is possible that the observed ring structure in the satellite galaxy could likely be due to the interaction event with the main galaxy. The tidal bridge situated between the pair exhibits moderate  $\mathrm{log[{\Sigma_{SFR}}_{mean}]} \sim -2.4$. {We also identified a star forming region on the northwest of NGC 4017, identified as \textit{Region I} in Figure \ref{fig:305ssfr} (left panel)}. We observe that the region is separated from the main disk of NGC 4017 and exhibits $\mathrm{log({\Sigma_{SFR}}_{mean} )} \sim -2.7$. These regions might have formed as a result of the interaction between the galaxies at the time of its closest approach \citep{toomre1972galactic,Hancock2009AJ....137.4643H}. Such features may arise due to the interaction between the pair. The orbital motion of gas slows down and causes a pile-up of gas, triggering enhanced star formation. We identify localised enhancements in star formation in addition to the disruption in morphology. We overplotted the H\textsc{i} column density map over the UV distribution in Figure \ref{fig:305ssfr} (right panel), with contours for column density values of $\mathrm{(0.8, 1,1.2,1.4) \times 10^{21} cm^{-2}}$. We observe that the regions with the $\mathrm{\Sigma_{SFR}}$ enhancement exhibit the highest H\textsc{i} column density values ($\mathrm{1.6 \times 10^{21} cm^{-2}}$). This observation aligns with our understanding that regions of intensified star formation are often associated with increased gas density. {We also observe that \textit{Region I} on the northwest of NGC 4017 appears to be embedded in an extension from the inner HI disk towards the region.} We suggest the region is possibly a young interaction debris. These observed disruptions in the H\textsc{i} distribution and formation of tidal tails and bridges are visible indicators of the effect of the interaction \citep{toomre1972galactic, Sengupta2017MNRAS.469.3629S}.

\begin{figure*}
\begin{center}
{\includegraphics[width = 1\textwidth]{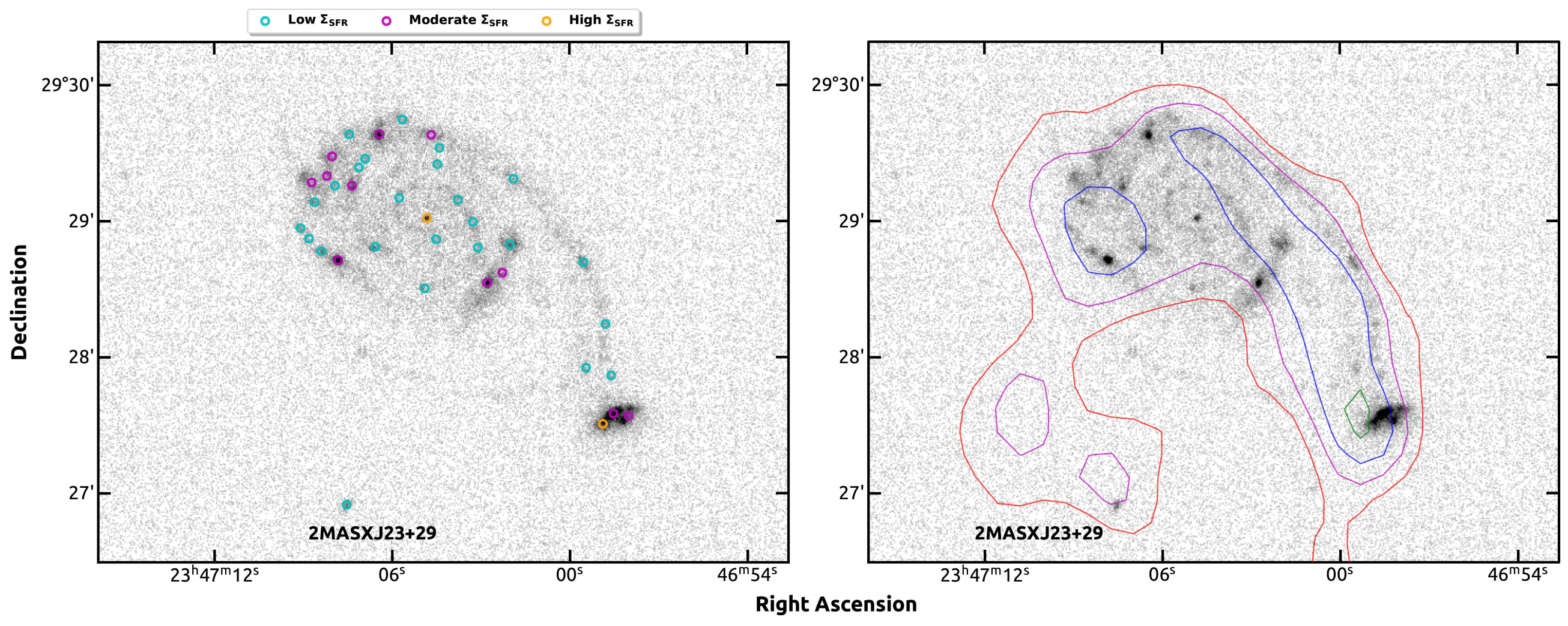}}
\caption{(Left) Spatial distribution of the $\mathrm{\Sigma_{SFR}}$ in the galaxy pair NGC 7753/7752 plotted on the FUV image. The TDG candidate \textit{2MASX J23470758+2926531}, denoted as \textit{2MASXJ23+29} in the image, exhibits moderate $\mathrm{\Sigma_{SFR}}$. (Right) H\textsc{i} column density contours are overplotted on the FUV image of NGC 7753/7752. The contours of red, magenta, blue and green colours represent H\textsc{i} column density values $\mathrm{(4, 6, 8, 10) \times 10^{20} cm^{-2}}$. \textit{2MASXJ23+29} is embedded in H\textsc{i} column density region $\mathrm{>1\times10^{21} cm^{-2}}$. }
\label{fig:86ssfr}
\end{center}
\end{figure*}

\subsubsection*{NGC 7753/7752} 

The distribution of $\mathrm{\Sigma_{SFR}}$ in the galaxy pair NGC 7753/7752 is depicted in Figure \ref{fig:86ssfr} (left panel). The identified regions in the galaxy pair exhibit $\mathrm{log({\Sigma_{SFR}})}$ in the range of -3 to -2. The central star-forming regions of the main galaxy exhibit the highest $\mathrm{log({\Sigma_{SFR}}_{mean} )} \mathrm{\sim -2}$. We observe that $\mathrm{\Sigma_{SFR}}$ is higher in the satellite galaxy than in the main galaxy. The regions identified in the satellite galaxy show an average value of $\mathrm{log({\Sigma_{SFR}}_{mean} )} \sim -2.18$. The satellite galaxy has regions with moderate to high $\mathrm{\Sigma_{SFR}}$. This can be understood better in light of the H\textsc{i} distribution. The bridge region extending from the main galaxy towards the satellite galaxy exhibits lower $\mathrm{\Sigma_{SFR}}$ values ($\mathrm{log[{\Sigma_{SFR}}_{mean}] \sim-2.84}$). \textit{2MASX J23470758+2926531}, the compact dwarf galaxy lying to the South-East of NGC 7752, shows only moderate star formation activity, with the mean $\mathrm{log({\Sigma_{SFR}}_{mean})\sim -2.6 }$.

We then study the distribution of H\textsc{i} in correlation with the UV distribution in Figure \ref{fig:86ssfr} (right panel). A possible reason for the observed high $\mathrm{\Sigma_{SFR}}$ in the satellite galaxy is the large H\textsc{i} content. The H\textsc{i} mass of the satellite galaxy is $\mathrm{{log(M_{H\textsc{i}}[M_\odot]) \sim 9.01}}$. The satellite galaxy exhibits a H\textsc{i} mass-to-total mass ratio of approximately $\sim0.17$ in comparison to typical starburst galaxies, which demonstrate a lower ratio of $\sim0.07$ \citep{Jackson_HIMASS1987AJ.....93..531J, Laurikainen1993ApJ...410..574L}. The derived H\textsc{i} ratio quantitatively measures the relative abundance of H\textsc{i} in the satellite galaxy with respect to its overall mass content. This implies that the satellite galaxy shows distinctive gas-rich characteristics compared to typical starburst galaxies, resulting in the observed enhancement of $\mathrm{\Sigma_{SFR}}$. We also observe that an H\textsc{i} bridge-like structure extends from the main galaxy towards the satellite galaxy and is embedded in a higher column density region $\mathrm{(\sim 10^{21} cm^{-2})}$. This provides evidence of the impact of galactic interaction on the restructuring of gas and thus, star formation in the galaxy. We also observe that \textit{2MASX J23470758+2926531} is embedded in the H\textsc{i} disk that extends from the main galaxy, confirming that the candidate is a probable member of the interacting system, and a possible candidate for TDG, formed as a result of the interaction episode. \\

On studying the spatial distribution of $\mathrm{\Sigma_{SFR}}$ in NGC 2207/IC 2163, NGC 4017/4016 and NGC 7753/7752 in light of the H\textsc{i} distribution, we understand that interaction events undergone by the sample of galaxies have triggered localised bursts of star formation, asymmetry in the distribution of H\textsc{i} and visible morphological distortions. {The localized variations in star formation activity and gas distribution are observed in galaxy regions affected by interaction events \citep{Smith2014AJ....147...60S,Smith2016AJ....151...63S}}. NGC 2207/IC 2163's near-gazing encounter manifests in a distributed enhancement of $\mathrm{\Sigma_{SFR}}$, correlated with varying H\textsc{i} column densities. Our identification of active star formation in \textit{N2207-NW1} may indicate towards the possibility of it being a possible TDG candidate. In NGC 4017/4016, we presence of enhanced star formation in the spiral arms of the galaxy, along with the asymmetry in H\textsc{i} distribution, highlights the enduring impact of the interaction event. In the case of NGC 7753/7752 pair, we observe a moderate to high $\mathrm{\Sigma_{SFR}}$ in the satellite galaxy, potentially attributed to its elevated H\textsc{i} content. We also observe the TDG \textit{2MASX J23470758+2926531}, which is embedded in H\textsc{i} contours that extends from the main galaxy exhibits moderate star formation activity. In addition to the localised star formation enhancements and distortion in the gas distribution, the interaction events may also affect the star formation properties on global scales \citep{Barrera2015A&A...582A..21B,Moreno2021MNRAS.503.3113M}. The extent to which these events may affect the galaxies is dependent on, but not limited to, a complex interplay between various factors, such as the stage of interaction, mass ratio of the interacting galaxies, separation between the galaxies etc. We probe these factors in the following section.

\subsection{Quantifying the effect of interaction on the main galaxies}\label{sec:knapen}

\begin{figure}
\begin{center}
{\includegraphics[width = 1\columnwidth]{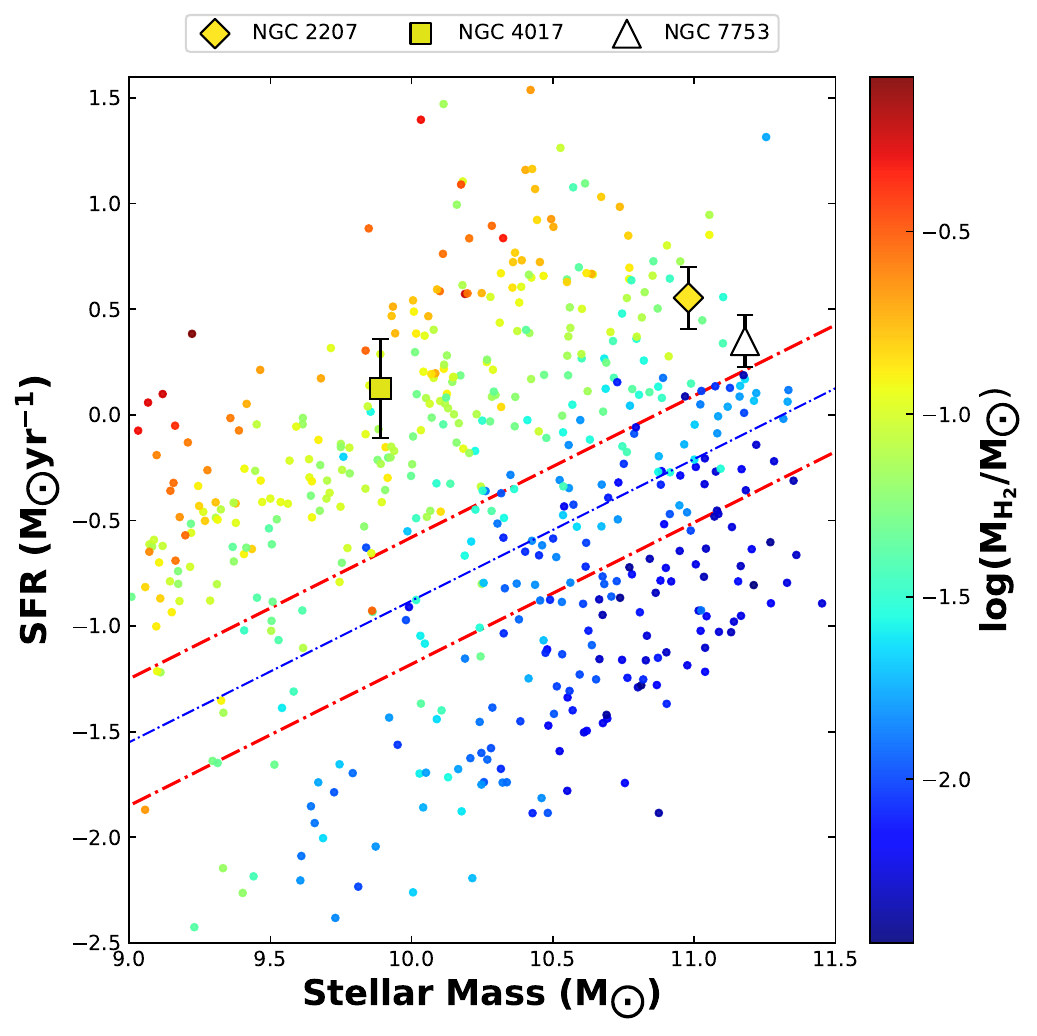}}
\caption{The SFMS of galaxies is presented. Circles represent the xCOLD GASS sample \citep{Saintonge2017ApJS..233...22S}, color mapped to represent the molecular gas fraction in them. The main sequence obtained from \citet{Dave2008MNRAS.385..147D} is represented by blue dotted lines, with the red dotted line representing 0.3 dex scatter. Our sample of main galaxies are also color mapped to represent the gas fraction. Due to the absence of relevant estimations of molecular gas mass in NGC 7753, its molecular gas fraction is not represented. We observe an enhancement in the global SFR in our sample of main galaxies.}
\label{fig:sfms}
\end{center}
\end{figure}

We understand that the interaction episodes can trigger localised star formation enhancements. To probe how these events may alter the gas distribution and star formation activity on global scales, we consider the global star formation activity in our sample of galaxies as a function of stellar mass. This distribution is known as the Star Forming Main Sequence (SFMS) and provides a baseline for understanding the star formation activity across galaxies of varying stellar content. We study the distribution of SFR in for the main galaxies as a function of the stellar mass. We consider the distribution observed in our sample by comparing it to the SFMS across galaxies studied in the xCOLD GASS sample \citep{Saintonge2017ApJS..233...22S}. \citet{Dave2008MNRAS.385..147D} considered the distribution and obtained a slope of $\mathrm{SFR \propto M_{\odot}^{0.77}}$ with 0.3 dex scatter for the main sequence. The position of galaxies in the SFMS is often a parameter of its molecular gas content. The magnitude of molecular gas fraction may determine the position of galaxy in the SFMS. Galaxies above the main sequence are typically known to be undergoing starbursts. Galaxies below the main sequence maybe experiencing episodes of star formation suppression \citep{Croton2006MNRAS.365...11C}. Thus, by considering the main sequence, we gain valuable insights into the conditions for star formation across the galaxy population and the physical processes that govern them. Figure \ref{fig:sfms} represents the obtained distribution, along with the main sequence obtained from \citet{Dave2008MNRAS.385..147D}. The galaxies studied by the xCOLD GASS sample are represented by closed circles and are color mapped to represent the molecular gas fraction. Our sample of main galaxies are also color mapped to represent the gas fraction. We obtained the molecular gas mass from \citet{Elmegreen2017ApJ...841...43E} for NGC 2207 and from \citet{Sengupta2017MNRAS.469.3629S} for NGC 4017. Due to the absence of relevant estimations of molecular gas mass in NGC 7753, we do not represent its molecular gas fraction. We observe an enhancement in the global SFR within our sample of main galaxies in Figure \ref{fig:sfms}. However, we need to consider that this observed enhancement likely arises due to influence from various parameters, potentially including merger classification, mass ratios of the interacting galaxies, and the separation distances between them. We probe the enhancement in global star formation activity in our sample of main galaxies as a function of these parameters.

We further study how these factors may affect star formation in interacting galaxies. The data sample from \citet{Knapen2015MNRAS.454.1742K} is used as the base for this analysis. The sample has almost 1478 galaxies within a distance of $\mathrm{\sim 60\ Mpc}$, of which 277 galaxies are interacting galaxies and the remaining, non-interacting systems. The non-interacting galaxy sample acts as a control sample for the study. The interacting galaxy sample is further sub-categorized based on the interaction class as (A) mergers, (B) highly distorted galaxies, (C) galaxies with minor distortions, and (O) galaxies with close companions \citep{Knapen2014A&A...569A..91K}. We classified the main galaxies in our sample, in addition to NGC 1512 from Paper I into their corresponding interaction class and the morphological type and defined control samples for each galaxy based on the above-discussed criterion. 

\citet{Knapen2015MNRAS.454.1742K} studied star formation in interacting galaxies in the IR regime. They compared the trends of star formation rate both in IR and UV and found slight differences between both. We create a superset of interacting galaxies that consists of the 277 interacting galaxies from \citet{Knapen2015MNRAS.454.1742K} and four galaxies from our sample. We study the correlation between the star formation and different physical parameters. 

\begin{figure}
\begin{center}
{\includegraphics[width = 1\columnwidth]{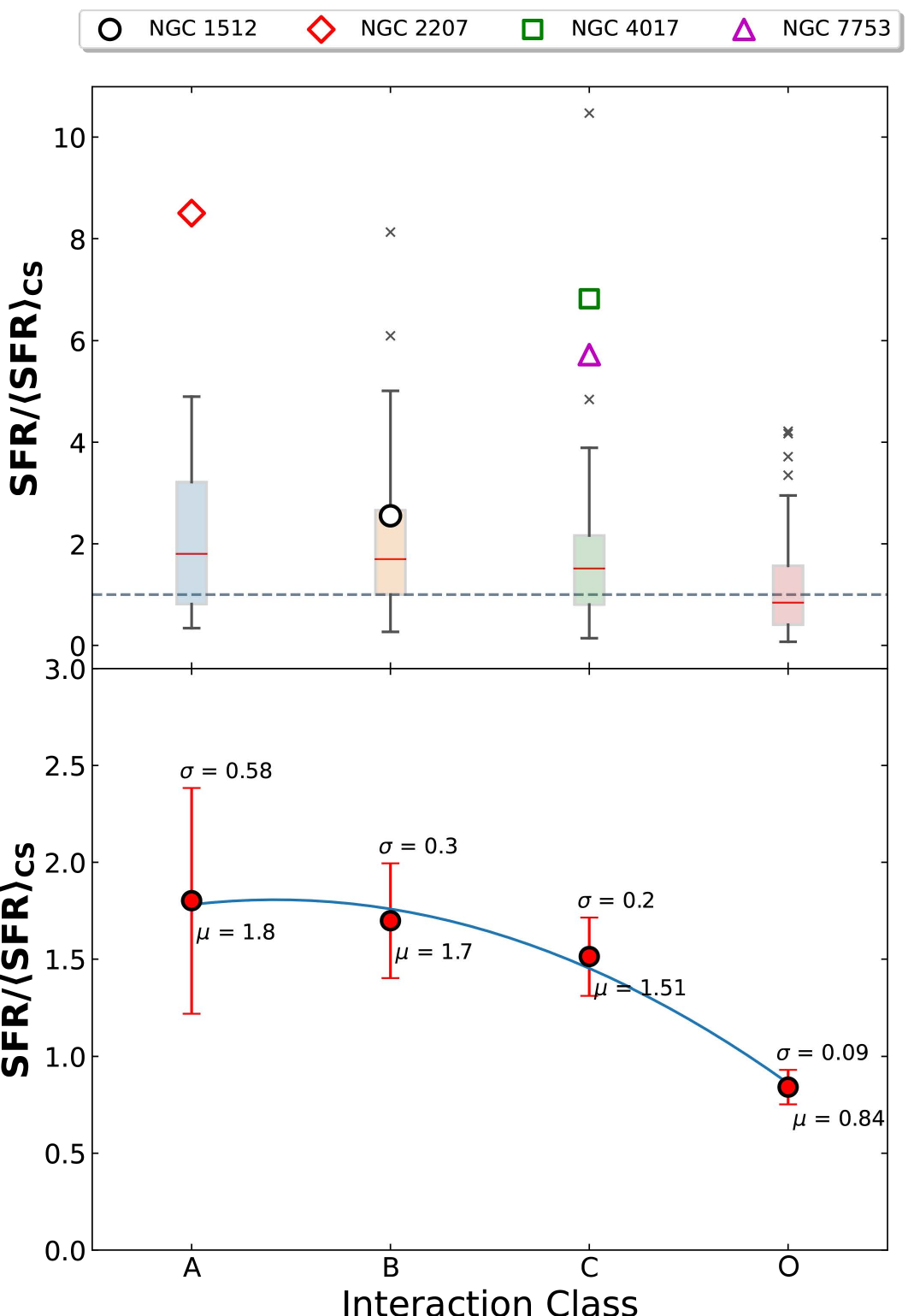}}
\caption{(Top) SFR enhancement, defined as the SFR of a galaxy normalized by the median SFR for its control population (see Section \ref{sec:knapen1}), separated by interaction class (where A is the class of mergers, and O contains those galaxies which are not interacting with their companions). {The boxplot represents the total spread of SFR enhancement for each interaction class.}  (Bottom) Median values per class are indicated with their $1\sigma$ uncertainty. A moderate enhancement exists in the median value of SFR enhancement across the interaction classes, as is evident from the distribution of median values. The median and $\sigma$ values are indicated next to the data points.}
\label{fig:sfrvsclass}
\end{center}
\end{figure}

\subsubsection{Studying the star formation enhancement as a function of interaction class}\label{sec:knapen1}

We define the enhancement in SFR as the SFR of the selected galaxy normalized to the median SFR of the control sample, $\mathrm{<SFR>_{CS}}$ \citep{Knapen2015MNRAS.454.1742K}. We studied the estimated enhancement in SFR as a function of the interaction classes A, B, C and O. The SFR in the various classes of galaxies are plotted for the sample and is given in Figure \ref{fig:sfrvsclass} (top panel). {The median values of SFR enhancement for each class with their $1\sigma$ is given in Figure \ref{fig:sfrvsclass} (bottom panel).} Though there is a huge spread in the SFR enhancement values within each interaction class, a moderate enhancement exists in the median value of SFR enhancement across the interaction classes, as is evident from the distribution of median values for each interaction class in Figure \ref{fig:sfrvsclass} (lower panel). However, a careful interpretation is essential considering the observed trends. First, there is a substantial dispersion in SFR enhancement within each interaction class, indicating a considerable overlap in enhancement among these classes. Second, the median level of SFR enhancement attributed to interactions remains moderate, with the most pronounced increase observed in the merger class, with a median of 1.8. We also observe in Figure \ref{fig:sfrvsclass} (top panel) a huge spread in the boxplot within each interaction class, suggesting the existence of huge spread of galaxies within each interaction class. A subset of galaxies displays SFR enhancement below unity, indicating their SFR is lower than those of the control sample galaxies. Another subset falls within the 1$\sigma$ range of the median value of SFR enhancement for Class A, while additional galaxies exhibit SFR enhancement values surpassing the 1$\sigma$ deviation from the median value. This observation suggests a huge spread among interacting galaxies, each characterized by distinct SFR enhancement.

\begin{figure*}
\begin{center}
{\includegraphics[width = 1\textwidth]{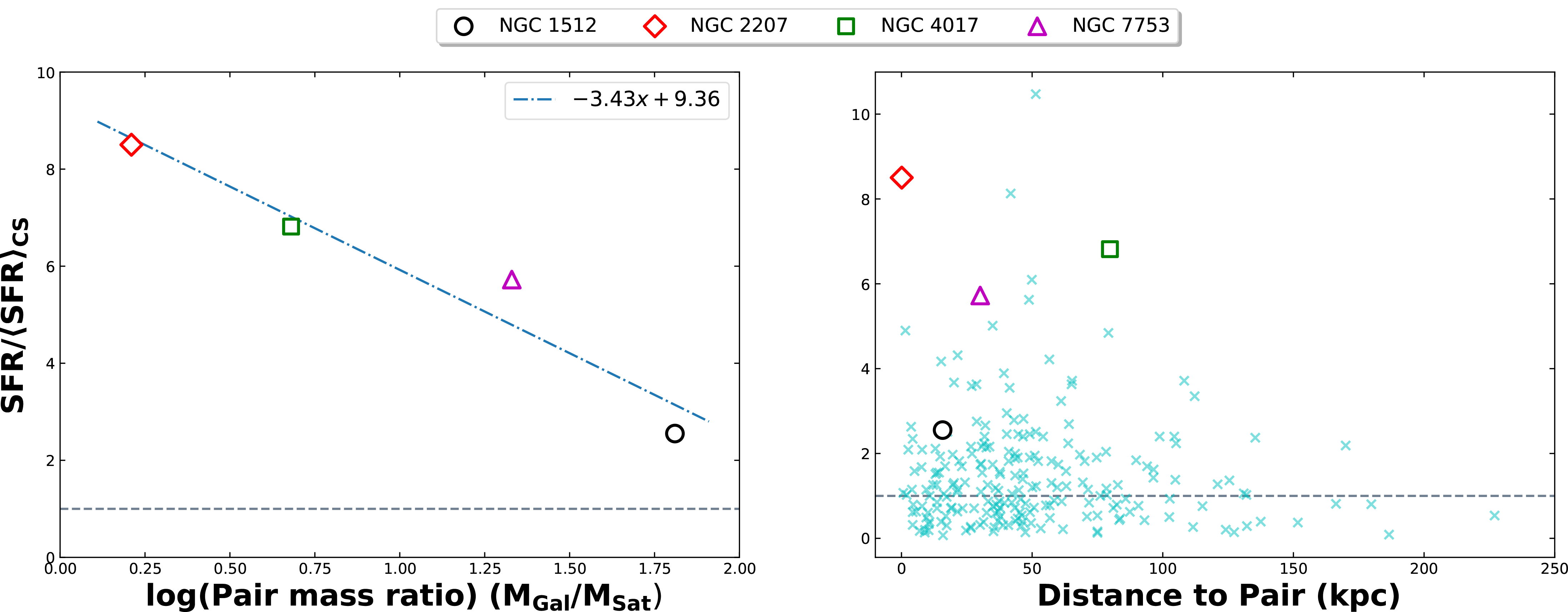}}
\caption{(Left) SFR enhancement studied as a function of the pair mass ratios. We find an inverse correlation between the SFR enhancement and mass ratios. (Right) SFR enhancement was studied as a function of the pair separation. Star formation enhancement for interacting galaxies from \citet{Knapen2015MNRAS.454.1742K} are also plotted as a function of pair separation. We find an absence of correlation, with a Pearson coefficient of 0.05. }
\label{fig:massrat}
\end{center}
\end{figure*}

These observations emphasize the complex nature of SFR enhancement resulting from galactic interactions and underscore the need for careful consideration in further analysis. The moderate enhancement, with the maximum of $\sim$1.8 in the merger class, might be due to the large-scale inflow of cold gas being driven by local instabilities triggered by the interaction episode. This supports our previous consideration that the `enhancement of SFR due to interaction' is only a broad assertion.  Our conclusions support previous simulation studies by \citet{Kapferer2005A&A...438...87K,Matteo2007A&A...468...61D,Matteo2008A&A...492...31D,Knapen2015MNRAS.454.1742K} and \citet{garcia2020A&A...635A.197D} that concluded only a moderate SFR enhancement in galaxies as due to interaction events. The enhancement in SFR is subjective to different parameters, such as the pair mass ratio and separation between the galaxies. Our investigation systematically examines the relationship between SFR enhancement and parameters of pair mass ratio and pair separation, shedding light on the complex relationship between galactic interactions and star formation processes. 

\subsubsection{Star formation enhancement as a function of pair mass ratio and as a function of pair separation}

We studied the enhancement of SFR in our sample of four interacting galaxy pairs as a function of the pair mass ratio. Pair mass ratio is defined as the ratio of the stellar mass of the main galaxy to that of the satellite galaxy ($\mathrm{M_{Gal}/M_{Sat}}$). The masses of the pair members were collated from Table \ref{tab:parameter}. 

We plotted the observed enhancement as a function of the pair mass ratio in Figure \ref{fig:massrat} (left panel). We observe an inverse correlation between the mass ratio of galactic pairs and SFR enhancement. Specifically, the observed enhancement in SFR resulting from galactic interactions is contingent upon the mass ratio of the interacting pair, with little to no enhancement detected in pairs characterized by high mass ratios. Interaction between galaxy pairs with equal mass members might generate stronger tidal forces, inducing stronger bursts of star formation. In cases where the interacting pairs have a higher $\mathrm{M_{Gal}/M_{Sat}}$ ratio, the SFR enhancement observed is little. This provides a correlation between the interaction-driven star formation enhancement and decreasing pair mass ratio in interacting galaxy pairs. Previous studies using numerical simulations observed that the merger-driven star formation is highly dependent on the pair mass ratio, with decreasing levels of interaction-induced star formation for increasing pair mass ratios \citep{Cox2008MNRAS.384..386C,Hani2020MNRAS.493.3716H}. Our present study provides observational evidence supporting the previously established relationship between SFR enhancement and pair mass ratio.

We also studied the enhancement as a function of the pair separation between the galaxies. Previous studies have probed the dependence of star formation enhancement on pair separation \citep{Nikolic2004MNRAS.355..874N, Woods2006AJ....132..197W,Patton2013MNRAS.433L..59P}. However, there is an absence of a clear understanding on the same. For our analysis, we studied the enhancement of SFR as a function of the pair separation. We plotted the SFR enhancement as a function of the pair separation, $\mathrm{R_{sep}}$, in Figure \ref{fig:massrat} (right panel). The dotted horizontal line denotes an SFR enhancement of unity. Galaxies above this have experienced an increase in SFR following an interaction event, while those below have not experienced a significant enhancement \citep{Knapen2015MNRAS.454.1742K}. We do not observe a correlation between the SFR enhancement and $\mathrm{R_{sep}}$. We examined this distribution in the sample of interacting galaxies from \citet{Knapen2015MNRAS.454.1742K} to explore such a correlation across a broader dataset. We found no significant correlation between the SFR enhancement and $\mathrm{R_{sep}}$, as evidenced by a Pearson coefficient of 0.05. Interestingly, while the galaxies NGC 1512 and NGC 7753 are closer to their satellite galaxies than NGC 4017, they exhibit lesser SFR enhancement than NGC 4017. This suggests that the interaction-induced star formation might be governed by additional factors beyond $\mathrm{R_{sep}}$. We consider the absence of correlation in light of their pair mass ratios. The lesser SFR enhancements observed in NGC 1512 and NGC 7753 might be due to their higher pair mass ratios than NGC 4017. This suggests that the dependence of SFR enhancement on pair separation must be considered in light of the pair mass ratio. We also understand that though there is an absence of a linear trend between pair separation and SFR enhancement, the pair separation can possibly act as a limiting parameter for the SFR enhancement, as inferred from Figure \ref{fig:massrat} (right panel). 

Our analysis of interacting pairs of galaxies indicates that the SFR enhancement is not solely determined by the separation distance between the interacting galaxies but also by the mass ratio of the galaxy pair. This finding highlights the importance of considering the pair mass ratio as a vital parameter when studying the SFR enhancement due to galaxy interactions. We conclude that SFR enhancement may be more pronounced when galaxies with comparable masses interact, as opposed to when galactic pairs with large mass ratios interact. We also understand that though there is no clear linear relation between the pair separation and SFR enhancement, the separation can act as a limiting parameters. There are other factors that may govern the SFR enhancement beyond the physical factors that we probed, such as, the velocity of approach, the geometry of the orbits in which the galaxies approach etc. Further studies will help us to understand the effect of such parameters on SFR enhancement. 

\section{Summary} \label{discussion}

Interactions between galaxies have long been recognized as one of the drivers of galactic evolution. The intense gravitational forces generated during these close encounters and mergers can trigger significant changes in the star formation rates, gas content, morphology, and overall structure of galaxies. Therefore, understanding the effects of interactions on galactic evolution is essential to develop a comprehensive picture of the formation and evolution of galaxies. 
In this study, we carried out an investigation on the star formation properties of three interacting galaxy pairs, namely NGC 2207/IC 2163, NGC 4017/4016 and NGC 7753/7752. Our aim was to investigate the recent star formation activity in these interacting galaxy pairs. Studying such interacting systems offers us a unique opportunity to comprehend the possible effects of such interaction events. Furthermore, it also enables us to understand better the extent to which parameters of pair mass ratio and pair separation might influence the impact of an interaction event on the involved galaxy pairs.

\begin{itemize}
    \item Based on the availability of UVIT data, we selected a sample of three interacting pairs of galaxies in field environments to study the effect of interactions on the star formation properties and subsequent galactic evolution.

    \item We studied star formation properties in the studied sample of galaxies. We observed localized enhancements of star formation in the galaxy pairs, possibly due to the interaction events that the galaxy pairs have undergone. We identified ongoing star-formation in probable TDG candidates in our sample of galaxies. We studied the UV distribution as a function of H\textsc{i} column density and found distortions in the regions of interaction. The probable TDG candidates are embedded in high H\textsc{i} regions.

    \item We probed the global star formation properties of the sample of galaxies in SFMS and identified them to lie above the main-sequence, pointing towards the enhancement action of the interaction event on a global scale.
    
    \item We studied the sample of galaxies with the inclusion of a bigger control sample to probe the factors affecting this enhancement action. We studied the SFR enhancement as a function of interaction class and found that the galaxies in the merger state experience the highest SFR enhancement of 1.8. The overall SFR enhancement is only moderate across the interaction classes.

    \item We studied the observed enhancement as a function of pair mass ratios. We find that the pair mass ratio is a dominant factor in deciding the enhancement in SFR, with galaxy pairs having lesser pair mass ratios experiencing greater enhancements than galaxy pairs for which pair mass ratio is much greater than 1. 
    
    \item We studied the observed enhancement as a function of pair separation and did not observe any correlation between the SFR enhancement and the separation distance. We observed that even though NGC 1512 and NGC 7753 are closer to their satellite galaxies than NGC 4017, they exhibit less SFR enhancement due to higher pair mass ratios. 
    
    \item This suggests that the dependence of SFR enhancement on pair separation must be considered in light of the pair mass ratio. We also infer that the pair separation can possibly act as a limiting parameter for the SFR enhancement.
    
\end{itemize}
Our findings provide new insights into the complex interplay between different physical parameters and the enhancement in SFR, shedding light on the mechanisms that drive the evolution of galaxies. This study based on a representative sample of nearby galaxies can be used to place stronger constraints on studies based on a wider sample of interacting galaxies.

\section*{Acknowledgements}

We thank the anonymous referee for the valuable comments and suggestions that have significantly improved the quality of the paper.  We would like to acknowledge the financial support from the Indian Space Research Organisation (ISRO) under the \textit{AstroSat} archival data utilization program (No. DS-2B-13013(2)/6/2019). We acknowledge the facility support from the FIST program of DST (SR/FST/PS-I/2022/208). AKR and SSK acknowledge the financial support from CHRIST (Deemed to be University, Bangalore) through the SEED money project (No: SMSS-2220, 12/2022). This publication uses the data from the UVIT, which is part of the \textit{AstroSat} mission of the ISRO, archived at the Indian Space Science Data Centre (ISSDC).We gratefully thank all the individuals involved in the various teams for providing their support to the project from the early stages of the design to launch and observations with it in the orbit. We thank the Center for Research, CHRIST (Deemed to be University) for all their support during the course of this work.

\section*{Data Availability}
The data underlying this article will be shared on reasonable request to the corresponding author.




\bibliographystyle{mnras}
\bibliography{ref} 

\appendix

\section{Distribution of $\mathrm{\Sigma_{SFR}}$ in the galaxy sample }

\begin{figure}
\begin{center}
{\includegraphics[width = 1\columnwidth]{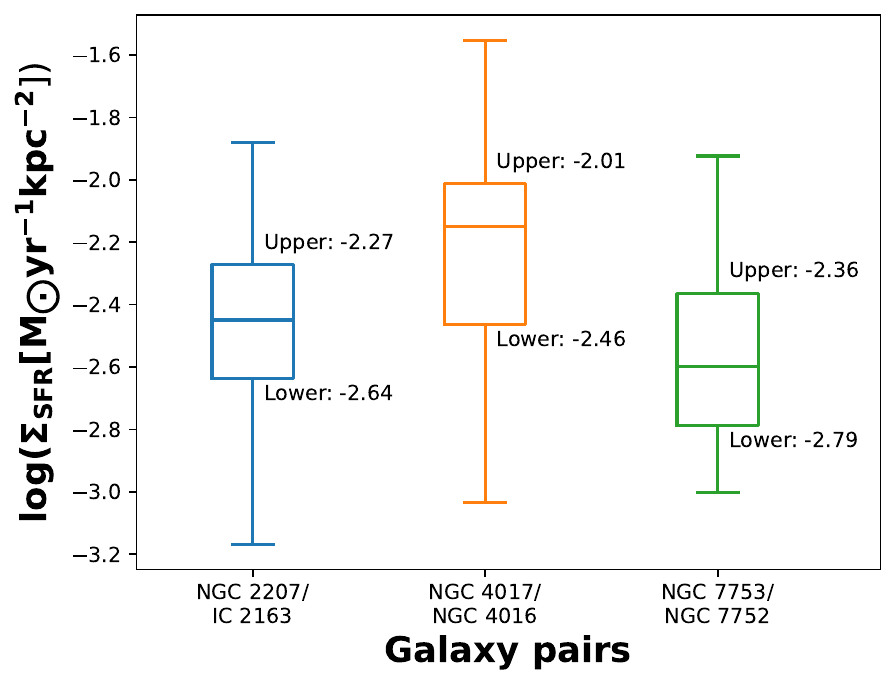}}
\caption{The range of $\mathrm{\Sigma_{SFR}}$ and the values of upper and lower quartiles for each galaxy in the selected sample. Regions with $\mathrm{\Sigma_{SFR}}$ falling above the upper quartile are identified as high $\mathrm{\Sigma_{SFR}}$ regions. Regions with $\mathrm{\Sigma_{SFR}}$ below the lower quartile are identified as low $\mathrm{\Sigma_{SFR}}$ regions while those exhibiting $\mathrm{\Sigma_{SFR}}$ between the upper and lower quartiles are identified as moderate $\mathrm{\Sigma_{SFR}}$ regions.}
\label{fig:append}
\end{center}
\end{figure}

\bsp	
\label{lastpage}
\end{document}